\input harvmac
%\draftmode
\let\includefigures=\iftrue
\let\useblackboard=\iftrue
\newfam\black 

%Figure Stuff
\includefigures
\message{If you do not have epsf.tex (to include figures),}
\message{change the option at the top of the tex file.}
\input epsf
\def\figin{\epsfcheck\figin}\def\figins{\epsfcheck\figins}
\def\epsfcheck{\ifx\epsfbox\UnDeFiNeD
\message{(NO epsf.tex, FIGURES WILL BE IGNORED)}
\gdef\figin##1{\vskip2in}\gdef\figins##1{\hskip.5in}% blank space instead
\else\message{(FIGURES WILL BE INCLUDED)}%
\gdef\figin##1{##1}\gdef\figinbs##1{##1}\fi}
\def\DefWarn#1{}
\def\figinsert{\goodbreak\midinsert}
\def\ifig#1#2#3{\DefWarn#1\xdef#1{fig.~\the\figno}
\writedef{#1\leftbracket fig.\noexpand~\the\figno}%
\figinsert\figin{\centerline{#3}}\medskip\centerline{\vbox{
\baselineskip12pt\advance\hsize by -1truein
\noindent\footnotefont{\bf Fig.~\the\figno:} #2}}
%\bigskip
\endinsert\global\advance\figno by1}
%%%
\else
\def\ifig#1#2#3{\xdef#1{fig.~\the\figno}
\writedef{#1\leftbracket fig.\noexpand~\the\figno}%
%\figinsert\figin{\centerline{#3}}\medskip
%\centerline{\vbox{\baselineskip12pt
%\advance\hsize by -1truein\noindent
%\footnotefont{\bf Fig.~\the\figno:} #2}}
%\bigskip\endinsert
\global\advance\figno by1} \fi

\def\id{{1 \kern-.28em {\rm l}}}

\def\K3{{\bf K3}}
\def\journal#1&#2(#3){\unskip, \sl #1\ \bf #2 \rm(19#3) }
\def\andjournal#1&#2(#3){\sl #1~\bf #2 \rm (19#3) }

\def\bar{\overline}

\def\eg{{\it e.g.}}

\def\tilde{\widetilde}

\def\frac#1#2{{#1\over#2}}
\def\coeff#1#2{{\textstyle{#1\over #2}}}

\def\inbar{\,\vrule height1.5ex width.4pt depth0pt}
\def\IC{\relax\hbox{$\inbar\kern-.3em{\rm C}$}}
\def\IR{\relax{\rm I\kern-.18em R}}
\def\IZ{\relax{\rm I\kern-.18em Z}}

%
%%%%%%%%%%%%%%%%%%%%%%%%%%%%%%%%%%%%
%

%
\catcode`\@=11
\def\slash#1{\mathord{\mathpalette\c@ncel{#1}}}
\overfullrule=0pt

\def\HH{{\cal H}}
\def\II{{\cal I}}

\def\NN{{\cal N}}

\def\RR{{\cal R}}

\def\WW{{\cal W}}

\def\underrel#1\over#2{\mathrel{\mathop{\kern\z@#1}\limits_{#2}}}

\catcode`\@=12

%%%%%%%%%%%%%%%%%%%%%%%%%%%%%%%%%%%%%%%%%%%%%%%%%%%%%%%%%%%%%%

%

\def\exp{{\rm exp}}

%%%%%%%%%%%%%%%%%%%%%%%%%%%%%%%%%%%%%%%%%%%%%%%%%%%%%%%%%%%%%%
% new defs:

\def\eg{{\it e.g.}}

%\SeibergPQ
\lref\SeibergPQ{
  N.~Seiberg,
  ``Electric - magnetic duality in supersymmetric nonAbelian gauge theories,''
Nucl.\ Phys.\ B {\bf 435}, 129 (1995).
[hep-th/9411149].
%%CITATION = hep-th/9411149%%
}

%\RomelsbergerEG
\lref\RomelsbergerEG{
  C.~Romelsberger,
  ``Counting chiral primaries in N = 1, d=4 superconformal field theories,''
Nucl.\ Phys.\ B {\bf 747}, 329 (2006).
[hep-th/0510060].
%%CITATION = hep-th/0510060%%
}

%\GaddeEN
\lref\GaddeEN{
  A.~Gadde, L.~Rastelli, S.~S.~Razamat and W.~Yan,
  ``On the Superconformal Index of N=1 IR Fixed Points: A Holographic Check,''
JHEP {\bf 1103}, 041 (2011).
[arXiv:1011.5278 [hep-th]].
%%CITATION = arXiv:1011.5278%%
}

%\RomelsbergerEC
\lref\RomelsbergerEC{
  C.~Romelsberger,
  ``Calculating the Superconformal Index and Seiberg Duality,''
[arXiv:0707.3702 [hep-th]].
%%CITATION = arXiv:0707.3702%%
}

%\KinneyEJ
\lref\KinneyEJ{
  J.~Kinney, J.~M.~Maldacena, S.~Minwalla and S.~Raju,
  ``An Index for 4 dimensional super conformal theories,''
Commun.\ Math.\ Phys.\  {\bf 275}, 209 (2007).
[hep-th/0510251].
%%CITATION = hep-th/0510251%%
}

\lref\KPS{
  D.~Kutasov, A.~Parnachev and D.~A.~Sahakyan,
  ``Central charges and U(1)(R) symmetries in N=1 superYang-Mills,''
JHEP {\bf 0311}, 013 (2003).
[hep-th/0308071].
%%CITATION = hep-th/0308071%%
}

%\BenvenutiQR
\lref\BenvenutiQR{
  S.~Benvenuti, B.~Feng, A.~Hanany and Y.~-H.~He,
  ``Counting BPS Operators in Gauge Theories: Quivers, Syzygies and Plethystics,''
JHEP {\bf 0711}, 050 (2007).
[hep-th/0608050].
%%CITATION = hep-th/0608050%%
}

%\FengUR
\lref\FengUR{
  B.~Feng, A.~Hanany and Y.~-H.~He,
  ``Counting gauge invariants: The Plethystic program,''
JHEP {\bf 0703}, 090 (2007).
[hep-th/0701063].
%%CITATION = hep-th/0701063%%
}

%\SpiridonovZA
\lref\SpiridonovZA{
  V.~P.~Spiridonov and G.~S.~Vartanov,
  ``Elliptic Hypergeometry of Supersymmetric Dualities,''
Commun.\ Math.\ Phys.\  {\bf 304}, 797 (2011).
[arXiv:0910.5944 [hep-th]].
%%CITATION = arXiv:0910.5944%%
}

%\DolanQI
\lref\DolanQI{
  F.~A.~Dolan and H.~Osborn,
  ``Applications of the Superconformal Index for Protected Operators and q-Hypergeometric Identities to N=1 Dual Theories,''
Nucl.\ Phys.\ B {\bf 818}, 137 (2009).
[arXiv:0801.4947 [hep-th]].
%%CITATION = arXiv:0801.4947%%
}

%\KutasovNP
\lref\KutasovNP{
  D.~Kutasov and A.~Schwimmer,
  ``On duality in supersymmetric Yang-Mills theory,''
Phys.\ Lett.\ B {\bf 354}, 315 (1995).
[hep-th/9505004].
%%CITATION = hep-th/9505004%%
}

%\KutasovSS
\lref\KutasovSS{
  D.~Kutasov, A.~Schwimmer and N.~Seiberg,
  ``Chiral rings, singularity theory and electric - magnetic duality,''
Nucl.\ Phys.\ B {\bf 459}, 455 (1996).
[hep-th/9510222].
%%CITATION = hep-th/9510222%%
}

%\KutasovVE
\lref\KutasovVE{
  D.~Kutasov,
  ``A Comment on duality in N=1 supersymmetric nonAbelian gauge theories,''
Phys.\ Lett.\ B {\bf 351}, 230 (1995).
[hep-th/9503086].
%%CITATION = hep-th/9503086%%
}

%\BrodieVX
\lref\BrodieVX{
  J.~H.~Brodie,
  ``Duality in supersymmetric SU(N(c)) gauge theory with two adjoint chiral superfields,''
Nucl.\ Phys.\ B {\bf 478}, 123 (1996).
[hep-th/9605232].
%%CITATION = hep-th/9605232%%
}

%\GaddeKB
\lref\GaddeKB{
  A.~Gadde, E.~Pomoni, L.~Rastelli and S.~S.~Razamat,
  ``S-duality and 2d Topological QFT,''
JHEP {\bf 1003}, 032 (2010).
[arXiv:0910.2225 [hep-th]].
%%CITATION = arXiv:0910.2225%%
}

%\IntriligatorMI
\lref\IntriligatorMI{
  K.~A.~Intriligator and B.~Wecht,
  ``RG fixed points and flows in SQCD with adjoints,''
Nucl.\ Phys.\ B {\bf 677}, 223 (2004).
[hep-th/0309201].
%%CITATION = hep-th/0309201%%
}

%\KutasovYQA
\lref\KutasovYQA{
  D.~Kutasov and J.~Lin,
  ``Exceptional N=1 Duality,''
[arXiv:1401.4168 [hep-th]].
%%CITATION = arXiv:1401.4168%%
}

%\KapustinVZ
\lref\KapustinVZ{
  A.~Kapustin, H.~Kim and J.~Park,
  ``Dualities for 3d Theories with Tensor Matter,''
JHEP {\bf 1112}, 087 (2011).
[arXiv:1110.2547 [hep-th]].
%%CITATION = arXiv:1110.2547%%
}

%\BashkirovVY
\lref\BashkirovVY{
  D.~Bashkirov,
  ``Aharony duality and monopole operators in three dimensions,''
[arXiv:1106.4110 [hep-th]].
%%CITATION = arXiv:1106.4110%%
}

%\ZwiebelWA
\lref\ZwiebelWA{
  B.~I.~Zwiebel,
  ``Charging the Superconformal Index,''
JHEP {\bf 1201}, 116 (2012).
[arXiv:1111.1773 [hep-th]].
%%CITATION = arXiv:1111.1773%%
}

%\AharonySX
\lref\AharonySX{
  O.~Aharony, J.~Marsano, S.~Minwalla, K.~Papadodimas and M.~Van Raamsdonk,
  ``The Hagedorn - deconfinement phase transition in weakly coupled large N gauge theories,''
Adv.\ Theor.\ Math.\ Phys.\  {\bf 8}, 603 (2004).
[hep-th/0310285].
%%CITATION = hep-th/0310285%%
}

%%%%%%%%%%%%%%%%%%%%%%%%%%%%%%%%%%%%%%%%%%%%%%%%%
\Title{}
{\vbox{\centerline{N=1 Duality and the Superconformal Index}
\bigskip
%\centerline{..}
}}
\bigskip

\centerline{\it David Kutasov and Jennifer Lin}
\bigskip
%\smallskip
\centerline{EFI and Department of Physics, University of
Chicago} \centerline{5640 S. Ellis Av., Chicago, IL 60637, USA }
\smallskip

\vglue .3cm

\bigskip

\let\includefigures=\iftrue
\bigskip
\noindent 
We use the $\NN=1$ superconformal index to study certain quantum constraints on chiral operators in a class of non-trivial SCFT's.

\bigskip

\Date{}

\newsec{Introduction}

Despite impressive progress in the last twenty years, our understanding of four dimensional $\NN=1$ supersymmetric field theory remains quite incomplete. A case in point is the series of fixed points discovered and studied in \refs{\KutasovVE\KutasovNP\KutasovSS\BrodieVX\KPS\IntriligatorMI-\KutasovYQA}. These models have gauge group $G=SU(N_c)$, $N_f$ chiral superfields in the fundamental representation of $G$ and two adjoint superfields $X$, $Y$. Their infrared physics is controlled by the superpotential for the adjoints $W(X,Y)$ and was shown in \IntriligatorMI\ to follow an ADE classification. 

The uniform construction of these models makes it natural to expect that they have many features in common. Indeed, in all the cases that have been understood to date, which include the $A_k$, $D_k$ and $E_7$ theories, one finds the same basic structure. The non-trivial fixed point exists when the number of flavors is in a  range, $N_{\rm min}\le N_f\le N_{\rm max}$, which depends on $N_c$. For $N_f<N_{\rm min}$ the theory does not have a supersymmetric vacuum, and the region just above $N_{\rm min}$, where the original theory is strongly coupled, has an alternative weakly coupled description, which utilizes  a  generalization of Seiberg duality \SeibergPQ.  

The $E_6$ and $E_8$ theories remain mysterious. As mentioned in \KutasovYQA, some aspects of the picture outlined above do not seem to generalize to them, and new elements are needed to understand their long distance physics. We will confirm the conclusions of \KutasovYQA\ from a different point of view. 

Even in the cases that are understood, there are aspects that remain puzzling. One involves the issue of quantum constraints on chiral operators. Such constraints occur in all the theories mentioned above, but they seem to play a more prominent role in the $D_k$ (with even $k$) and $E_7$ theories (see \refs{\BrodieVX,\IntriligatorMI,\KutasovYQA} for further discussion). The origin of these constraints is unclear, and one of the motivations of this work is to understand them better.

The tool we will use for this purpose is the superconformal index that has been developed in recent years \refs{\RomelsbergerEG\RomelsbergerEC-\KinneyEJ}. 
This index is a regularized Witten index for the theory on ${\bf R \times S^3}$ which counts short multiplets that cannot combine into long ones. For conformal field theories, ${\bf R \times S^3}$ and  {\bf R$^4$} are equivalent, since one can map one to the other by a conformal transformation. Under this transformation, 
the dilatation operator on ${\bf R^4}$ is mapped to the Hamiltonian (the generator of translations along ${\bf R}$) on ${\bf R \times S^3}$, and the $SO(4)\simeq SU(2)\times SU(2)$ rotation group of ${\bf R^4}$ is mapped to the isometry group of the ${\bf S^3}$. 

The $\NN=1$ supercharges $Q_{\alpha}$ and $\bar Q_{\dot\alpha}$ form doublets of the two $SU(2)$'s. Consider one of these supercharges, $\bar Q_+$, which carries charge $-1/2$ under the Cartan subalgebra generator of one of the two $SU(2)$'s, $\bar J_3$. The superconformal algebra includes the anti-commutation relation 
 \eqn\antic{
 \{\bar Q_+, \bar Q_+^\dagger\} =  \HH  = 2\left (H - 2 \bar J_3 - \frac 32 R \right)
}
where $H$ is the Hamiltonian and $R$ is the generator of the superconformal $U(1)_R$, normalized such that the $R$-charge of $\bar Q_+$ is $+1$. The operator $\HH$ commutes with the supercharges $(\bar Q_+,\, \bar Q_+^\dagger)$, and has the further property that all of its eigenvalues are non-negative.  It is therefore natural to define the Witten index 
\eqn\wi{
\II_W = \Tr(-1)^F e^{-\beta \HH}
}
with the trace taken over all states in the theory. Because states with $\HH > 0$ come in boson/fermion pairs whose contributions cancel, only states with $\HH=0$ contribute to \wi, which is therefore independent of $\beta$.

The number of states with $\HH=0$ is generically infinite,\foot{This is the case even in free field theory.} which makes the index \wi\ ill defined. To resolve the degeneracy, one can refine \wi\ by introducing chemical potentials for operators that commute among themselves and with $\bar Q_+, \bar Q_+^\dagger$. 
One such operator is
\eqn\defr{
\RR = R + 2\bar J_3 + \HH\,.
}
Another is the Cartan generator of the other $SU(2)$ in $SO(4)$, $J_3$. Introducing chemical potentials $t,x$ for them leads to the following generalization of the Witten index \RomelsbergerEC,
\eqn\defindex{
\II(t,x) = \Tr(-1)^F x^{2J_3}t^{\RR}e^{-\beta \HH}.
}
Further extensions are obtained by introducing chemical potentials for the Cartan subalgebra generators of the global symmetry group. A convenient way to impose
gauge invariance is to introduce chemical potentials for the gauge generators, and integrate over them.

The superconformal index has some properties that make it useful for studying strongly coupled $\NN=1$ SCFT's. In particular,  in a family of SCFTs parametrized by the coefficients of  exactly marginal operators, the index is invariant under changes of these parameters.  If the manifold of fixed points includes weakly coupled theories (as \eg\ in $\NN=4$ SYM), one can compute the index for them, and it must be the same in the strongly coupled regime. 

Another class of strongly coupled SCFT's for which weak coupling techniques can be used to evaluate the index are those connected to weakly coupled ones via RG flow \RomelsbergerEC. This is particularly clear in theories in which the $U(1)_R$ current that becomes part of the superconformal algebra in the IR is preserved throughout the RG flow, as is  the case \eg\ in the conformal windows of the ADE theories mentioned above. One can then study the theory on ${\bf R \times S^3}$ (which is of course not equivalent to that on {\bf R$^4$} at generic points along the RG flow), and compute the index \defindex, with $R$ the conserved $R$-charge. The index is independent of the RG parameter  $r\Lambda$ with $r$ the radius of ${\bf S}^3$ and $\Lambda$ the RG scale. Thus, we can compute it in the weakly coupled UV theory and it must agree with the superconformal index of the IR SCFT.

In many cases (\eg\ the ADE theories outside their conformal windows), the infrared superconformal $U(1)_R$ is not preserved throughout the RG flow, but rather is an accidental symmetry of the infrared theory. In those cases too we can compute the index \defindex, as long as there is an $R$-charge, $R$, that is conserved throughout the RG flow. Essentially by definition, the infrared superconformal $U(1)_R$ differs from $R$ by an accidental non-$R$ global symmetry.  We can think of the supersymmetric partition sum \defindex\ in these cases as the superconformal partition sum with a fugacity for this global symmetry that is correlated with that of the $U(1)_R$. For many applications (such as studying $\NN=1$ dualities) this correlation is not a problem. 

To compute the index for an $\NN=1$ SYM theory with gauge group $G$, flavor group $F$, and  chiral multiplets $\Phi_i$ in representations $R_{F,i}$ and $R_{G,i}$ of the flavor and gauge group, whose lowest components have R-charge $r_i$, we first evaluate the trace \defindex\ over single-particle states \RomelsbergerEC\
\eqn\spi{
i(t,x,z,y) = \frac{2t^2 - t(x + x^{-1})}{(1-tx)(1-tx^{-1})}\chi_{\rm adj_G}(z) + \sum_i \frac{t^{r_i}\chi_{R_F,i}(y)\chi_{R_G,i}(z) - t^{2 - r_i}\chi_{\bar R_F,i}(y) \chi_{\bar R_G, i}(z)}{(1-tx)(1-tx^{-1})}\,.
}
Here $\chi_{R}$ is the character of the representation $R$, and $\bar R$ denotes the conjugate representation. The arguments of the characters are eigenvalues of the group elements in the given representation $R$. \foot{
Given an element $g$ of a group $G$, the character $\chi_{R_G}(g) : G \rightarrow {\bf C}$ is defined to be the trace of $g$ in the representation $R$. 
One can write $g$ as a matrix of size $\dim R$, and diagonalize this matrix.
Then $\chi_R(g)$ is the sum of the eigenvalues of the matrix.
So we can take the argument of $\chi_R$ to be the $\dim R$ eigenvalues in the first place, instead of the group elements.
}
%
% fugacities. 
%
Reviews of group characters can be found in \refs{\AharonySX,\ZwiebelWA}.
 For later use, if we parametrize the $N-1$ independent eigenvalues of the fundamental of  $SU(N)$ as
\eqn\zzd{ z_i \in {\bf C},\,\, i \in 1, \dots, N; \quad \prod_i z_i = 1\,,}
then the characters of the representations that we will need are
\eqn\chs{\eqalign{
&\chi_{\rm fund_{SU(N)}}(z) = %p_N(z) =  
\sum_{i=1}^N z_i\,; \qquad \qquad
\chi_{\bar{\rm fund}_{SU(N)}}(z) = %p_N(z^{-1}) =
 \sum_{i=1}^N z_i^{-1}\,, \cr
&\chi_{\rm adj_{SU(N)}}(z)  =  %p_N(z)p_N(z^{-1}) -1 = 
\sum_{1 \leq i,j \leq N} z_i / z_j -1 =  \chi_{\rm fund_{SU(N)}}(z)\chi_{\bar{\rm fund}_{SU(N)}}(z)-1\,.
}}
The single-letter index \spi\ counts states with $\HH = 0$ \antic\ where the vacuum is acted on just once with a creation operator in the free field limit. Equivalently, it counts $\HH=0$ operators of the schematic form $\phi, \partial\phi, \dots, \partial^n\phi$ for arbitrary $n$ but excluding products of two or more $\phi$'s, with $\phi$ parametrizing all fields in the various multiplets, and $\partial$ all spacetime derivatives.

The first term in \spi\ is the contribution of the vector multiplet while the second comes from chiral multiplets. The terms in the second proportional to $t^{r_i}$ and $t^{2- r_i}$ are respectively from the scalar in the $i$th chiral multiplet with R-charge $r_i$, and its conjugate fermion partner with R-charge $1 - r_i$ and $\bar J_3$ charge $1/2$. Other components of the chiral multiplet have $\HH\neq 0$ and do not contribute to \spi. The denominator of \spi\ encodes the action of the spacetime derivatives $\partial_{\alpha\dot\alpha}= \partial_{++},\, \partial_{-+}$ which have $(H, \bar J_3, R) = (1,\frac 12, 0)$, giving $\HH=0$.

Given \spi, the full index is given by 
\eqn\fulli{
\II(t,x,y) = \int_G d\mu \,\exp\left(\sum_{n=1}^\infty \frac 1 n i(t^n, x^n, z^n, y^n) \right)
}
where $z^n$ is shorthand for taking each of the multiple $z_i$ (as \eg\ in \zzd)  to the $n$th power, and likewise for $y^n$. 
The integrand is the plethystic exponential of \spi, which returns all strings of operators that one can form by taking products of the operators contributing to \spi\ \refs{\BenvenutiQR,\FengUR}.
The integral over the gauge group, with $d\mu$ the Haar measure on the group manifold (whose explicit form for $SU(N)$, as a function of the $z$'s,  is in Appendix A),
projects onto gauge singlets. 

We now return to the question of whether one can use the index to shed light on the appearance of quantum constraints on the chiral ring of certain SQFT's. One can imagine two possible approaches. In principle, the constraints should be visible in the index by expanding it to the appropriate order in the fugacities and seeing if operators affected by the constraints contribute. In practice, we find that this is difficult, for reasons that will be explained later.

A second way that the index can shed light on such issues is by using the phenomenon of duality. Under $\NN=1$ duality, a theory with gauge group $SU(N_c)$ and $N_f$ flavors is equivalent in the IR to a magnetic theory with a different group $SU(\tilde N_c)$. The magnetic theory contains, in addition to fields that are charged under the gauge group, gauge singlet mesons that map to composite operators in the electric theory. As we will see, assuming that such a duality exists, by comparing the superconformal indices of the electric and magnetic theories, one can read off the spectrum of electric mesons, and in particular any quantum constraints that need to be imposed on them.  We will see that this approach is quite effective for the ADE theories.

Although the superconformal indices of theories with arbitrary gauge groups and matter can be written as elliptic hypergeometric integrals (see Appendix A), the group integrals are often difficult to evaluate. As such, demonstration of equality between the indices of dual theories often relies on complicated mathematical identities.\foot{The equality of the exact superconformal indices for electric and magnetic SQCD was proven in \DolanQI; the formulae for the indices of the $A_k$ and $D_k$ generalizations are written in \SpiridonovZA, but their equality remains a conjecture.} However in \DolanQI, Dolan and Osborn showed that the index simplifies greatly in the large $N$ Veneziano limit. 

Specifying to gauge group $SU(N)$, the single-letter index \spi\ can be put into the form
\eqn\spdo{
i(t,z) = f(t)(p_N(z)p_N(z^{-1})-1) + g(t) p_N(z)+ \bar g(t) p_N(z^{-1}) + h(t)
}
where to match to the notation of \DolanQI, we have renamed the $SU(N)$ characters \chs\ as
\eqn\pnotn{\eqalign{
&\chi_{\rm fund}(z) = p_N(z);  \qquad \chi_{\rm\bar{fund}}(z) = p_N(z^{-1});   \cr
&\chi_{\rm adj}(z) = p_N(z)p_N(z^{-1})-1\,,
}}
and $t$ in \spdo\ is shorthand for all other variables $(t,x,y)$ in \spi. 
Here $f(t), g(t), h(t)$ are defined implicitly by comparing \pnotn\ with \spi; we will give them explicitly in examples below.
 $f(t)$ is the contribution of fields in the adjoint of the gauge group (including the vector multiplet), $g(t)$ and $\bar g(t)$ are those of the fundamental and antifundamental superfields respectively, and $h(t)$ comes from gauge singlets.

Inserting \spdo\ into \fulli\ and using large $N$ orthogonality properties of power-symmetric polynomials (see \DolanQI\ for details), the full index \fulli\ in the Veneziano limit is given by
\eqn\iln{
\II(t) = \exp\left(\sum_{n=1}^\infty \frac 1 n \left(\frac{g(t^n)\bar g(t^n)}{1 - f(t^n)} - f(t^n) + h(t^n) \right) \right) \prod_{n=1}^\infty \frac{1}{1 - f(t^n)}\,.
}
In this limit, the comparison of electric and magnetic indices is easier. 

Our strategy below is to assume that each $SU(N_c)$ gauge theory in the aforementioned ADE classification has an $SU(\tilde N_c)$ dual description with $\alpha$ gauge singlet mesons that represent composite objects in the electric theory, in addition to the fundamental and adjoint matter which mirrors that of the electric theory. In other words, we assume the structure appearing in known dualities, but without specifying the meson spectrum. 't Hooft anomaly matching implies that $\tilde N_c$ must equal $\alpha N_f - N_c$ where $N_f$ is the number of flavors in the electric (and magnetic) theory \KutasovYQA. We will see that comparison of the large $N$ indices of the electric theory and its conjectured dual give powerful constraints on $\alpha$ and the R-charges of the mesons, that allows one to determine them, in cases where a duality of this form exists.

The outline of the paper is as follows. In section 2 we use the superconformal index to rederive the duality of \refs{\KutasovVE\KutasovNP-\KutasovSS} for the $A$ series models. In particular, we use the simplification of the index in the large $N$ Veneziano limit to obtain the spectrum of singlet mesons in the magnetic theory. In sections 3 and 4 we repeat the discussion for the $D$ and $E_7$ cases, using the index as a check on the appearance of the quantum constraints on the chiral ring mentioned above. In section 5 we briefly comment on the generalization of the discussion to the $E_6$ and $E_8$ cases. Two appendices contain some further useful properties of the index. 

\newsec{$A$ series}

In \DolanQI, Dolan and Osborn used \iln\ to verify that the superconformal indices of the dual descriptions for $\NN=1$ SQCD with one adjoint chiral superfield and a polynomial superpotential \refs{\KutasovVE\KutasovNP-\KutasovSS} agree in the Veneziano limit. This provided nontrivial evidence for the validity of the duality. 
The exact formula for the index of these theories, valid for all $N_c$, was also written down in \DolanQI, but it is not known whether it agrees in the electric and magnetic theories.  

As a warm-up, in this section we will revisit this calculation. However, we will perform it in reverse. Rather than checking that the large $N$ indices of the known electric and magnetic theories agree, we will assume that the electric theory is dual to an unknown theory with gauge group $SU(\alpha N_f - N_c)$ and  $\alpha$ gauge singlet mesons in the bifundamental of the flavor group. We will then use the equality of the indices to derive $\alpha$ and the R-charges of the mesons.

The electric $A_k$ theory has gauge group $SU(N_c)$, global symmetry group $SU(N_f) \times SU(N_f) \times U(1)_B\times U(1)_R$, $N_f$ flavors $Q, (\tilde Q)$ in the (anti)fundamental of the gauge group, and an adjoint matter field $X$ with superpotential 
\eqn\akw{\WW = \Tr X^{k+1}\,.
} 
The transformation properties of the various fields under the symmetries are summarized in the following table:
\bigskip
\vbox{
$$\vbox{\offinterlineskip
\hrule height 1.1pt
\halign{&\vrule width 1.1pt#
&\strut\quad#\hfil\quad&
\vrule width 1.1pt#
&\strut\quad#\hfil\quad&
\vrule width 1.1pt#
&\strut\quad#\hfil\quad&
\vrule width 1.1pt#
&\strut\quad#\hfil\quad&
\vrule width 1.1pt#
&\strut\quad#\hfil\quad&
\vrule width 1.1pt#
&\strut\quad#\hfil\quad&
\vrule width 1.1pt#\cr
height3pt
&\omit&
&\omit&
&\omit&
&\omit&
&\omit&
&\omit&
\cr
&\hfil Field&
&\hfil $SU(N_c)$&
&\hfil $SU(N_f)$&
&\hfil $SU(N_f)$&
&\hfil $U(1)_B$&
&\hfil $U(1)_R$&
\cr
height3pt
&\omit&
&\omit&
&\omit&
&\omit&
&\omit&
&\omit&
\cr
\noalign{\hrule height 1.1pt}
height3pt
&\omit&
&\omit&
&\omit&
&\omit&
&\omit&
&\omit&
\cr
&\hfil $Q$&
%&\hfil $N_c$& 
&\hfil $f$&
&\hfil $f$&
&\hfil $1$&
&\hfil $1$&
&\hfil $1 - \frac 2{k+1}\frac{N_c}{N_f}$& 
\cr
height3pt
&\omit&
&\omit&
&\omit&
&\omit&
&\omit&
&\omit&
\cr
%\noalign{\hrule}
%height3pt
&\omit&
&\omit&
&\omit&
&\omit&
&\omit&
&\omit&
\cr
&\hfil $\tilde Q$&
%&\hfil $N_c$& 
&\hfil $\bar f$&
&\hfil $1$&
&\hfil $\bar f$&
&\hfil $-1$&
&\hfil $1 - \frac 2{k+1}\frac{N_c}{N_f}$& 
\cr
%\noalign{\hrule}
height3pt
&\omit&
&\omit&
&\omit&
&\omit&
&\omit&
&\omit&
\cr
&\hfil $V$&
%&\hfil $\bar N_c$&
&\hfil ${\rm adj.}$&
&\hfil $1$&
&\hfil $1$&
&\hfil $0$&
&\hfil $0$&
\cr
%\noalign{\hrule}
height3pt
&\omit&
&\omit&
&\omit&
&\omit&
&\omit&
&\omit&
\cr
&\hfil $X$&
%&\hfil $\bar N_c$&
&\hfil ${\rm adj.}$&
&\hfil $1$&
&\hfil $1$&
&\hfil $0$&
&\hfil $\frac 2{k+1}$&
\cr
%\noalign{\hrule}
height3pt
&\omit&
&\omit&
&\omit&
&\omit&
&\omit&
&\omit&
\cr
}\hrule height 1.1pt
}
$$
}
\centerline{Table 1: The field content of the $A_k$ electric theory.} 
\smallskip
\noindent
In the following, we will make use of the parametrization
\eqn\defpq{
p = tx, \qquad q = tx^{-1}
}
and take $y, \tilde y, z$ to be the complex eigenvalues of the two $SU(N_f)$'s and $SU(N_c)$ respectively (as in \zzd). Also introducing $v$ as the chemical potential for  $U(1)_B$,
the single-particle index \spi\ of the theory is \DolanQI\
\eqn\aie{\eqalign{
& i_E(p,q,v, y, \tilde y, z)  \cr
&= -\left(\frac{p}{1-p} + \frac{q}{1-q} - \frac{1}{(1-p)(1-q)}((pq)^{\frac 12r_X} - (pq)^{1 - \frac 12 r_X}) \right)(p_{N_c}(z) p_{N_c}(z^{-1})-1) \cr
&\quad  + \frac{1}{(1-p)(1-q)}\left((pq)^{\frac 12 r_Q}vp_{N_f}(y)p_{N_c}(z) - (pq)^{1 - \frac 12 r_Q}v^{-1}p_{N_f}(y^{-1})p_{N_c}(z^{-1}) \right. \cr
&\qquad\qquad\qquad\qquad +  \left. (pq)^{\frac 12 r_Q}v^{-1}p_{N_f}(\tilde y^{-1}) p_{N_c}(z^{-1}) - (pq)^{1 - \frac 12 r_Q}v p_{N_f}(\tilde y)p_{N_c}(z) \right)\,,
}}
where $r_X, r_Q$ are the R-charges of the corresponding fields listed in Table 1.

We now assume that the $A_k$ theory has a magnetic dual description with gauge group $SU(\tilde N_c) = SU(\alpha N_f - N_c)$ and the field content
\bigskip
\vbox{
$$\vbox{\offinterlineskip
\hrule height 1.1pt
\halign{&\vrule width 1.1pt#
&\strut\quad#\hfil\quad&
\vrule width 1.1pt#
&\strut\quad#\hfil\quad&
\vrule width 1.1pt#
&\strut\quad#\hfil\quad&
\vrule width 1.1pt#
&\strut\quad#\hfil\quad&
\vrule width 1.1pt#
&\strut\quad#\hfil\quad&
\vrule width 1.1pt#
&\strut\quad#\hfil\quad&
\vrule width 1.1pt#\cr
height3pt
&\omit&
&\omit&
&\omit&
&\omit&
&\omit&
&\omit&
\cr
&\hfil Field&
&\hfil $SU(\tilde N_c)$&
&\hfil $SU(N_f)$&
&\hfil $SU(N_f)$&
&\hfil $U(1)_B$&
&\hfil $U(1)_R$&
\cr
height3pt
&\omit&
&\omit&
&\omit&
&\omit&
&\omit&
&\omit&
\cr
\noalign{\hrule height 1.1pt}
height3pt
&\omit&
&\omit&
&\omit&
&\omit&
&\omit&
&\omit&
\cr
&\hfil $q$&
%&\hfil $N_c$& 
&\hfil $f$&
&\hfil $\bar f$&
&\hfil $1$&
&\hfil $N_c/\tilde{N_c}$&
&\hfil $1 - \frac 2{k+1}\frac{\tilde N_c}{N_f}$& 
\cr
height3pt
&\omit&
&\omit&
&\omit&
&\omit&
&\omit&
&\omit&
\cr
%\noalign{\hrule}
%height3pt
&\omit&
&\omit&
&\omit&
&\omit&
&\omit&
&\omit&
\cr
&\hfil $\tilde q$&
%&\hfil $N_c$& 
&\hfil $\bar f$&
&\hfil $1$&
&\hfil $f$&
&\hfil $-N_c/\tilde{N_c}$&
&\hfil $1 - \frac 2{k+1}\frac{\tilde N_c}{N_f}$& 
\cr
%\noalign{\hrule}
height3pt
&\omit&
&\omit&
&\omit&
&\omit&
&\omit&
&\omit&
\cr
&\hfil $\tilde V$&
%&\hfil $\bar N_c$&
&\hfil ${\rm adj.}$&
&\hfil $1$&
&\hfil $1$&
&\hfil $0$&
&\hfil $0$&
\cr
%\noalign{\hrule}
height3pt
&\omit&
&\omit&
&\omit&
&\omit&
&\omit&
&\omit&
\cr
&\hfil $\tilde X$&
%&\hfil $\bar N_c$&
&\hfil ${\rm adj.}$&
&\hfil $1$&
&\hfil $1$&
&\hfil $0$&
&\hfil $\frac 2{k+1}$&
\cr
%\noalign{\hrule}
height3pt
&\omit&
&\omit&
&\omit&
&\omit&
&\omit&
&\omit&
\cr
&\hfil $M_j, \,\, j = 1, \dots \alpha$&
%&\hfil $\bar N_c$&
&\hfil $1$&
&\hfil $f$&
&\hfil $\bar f $&
&\hfil $0$&
&\hfil $2 r_Q + r_j$&
\cr
%\noalign{\hrule}
height3pt
&\omit&
&\omit&
&\omit&
&\omit&
&\omit&
&\omit&
\cr
}\hrule height 1.1pt
}
$$
}
\centerline{Table 2: The ``conjectured" field content of the $A_k$ magnetic theory.} 
\bigskip
\noindent
The $U(1)_R$ charges of the magnetic quarks are determined by anomaly considerations, as in table 1. The baryon charges are fixed by the baryon matching, $q^{\tilde N_c}\leftrightarrow Q^{N_c}$. The R-charge of $M_j$ is written as $2r_Q + r_j$, since this operator is mapped under duality to an electric operator of the form $\tilde Q\Theta_j Q$. The R-charges of the operators $\Theta_j$, $r_j$, are unspecified for now, and will be solved for in the following. 

The single-particle index of the theory in Table 2 is
\eqn\aim{\eqalign{
&i_M(p,q,v,y,\tilde y, \tilde z) \cr
&= -\left(\frac{p}{1-p} + \frac{q}{1-q} - \frac{1}{(1-p)(1-q)}((pq)^{\frac 12 r_X} - (pq)^{1 - \frac 12 r_X}) \right)(p_{\tilde N_c}(\tilde z) p_{\tilde N_c}(\tilde z^{-1})-1) \cr
&\quad + \frac{1}{(1-p)(1-q)}\left((pq)^{\frac 12 r_q} v^{N_c/\tilde N_c} p_{N_f}(y^{-1}) p_{\tilde N_c}(\tilde z) - (pq)^{1 - \frac 12 r_q} v^{-N_c/\tilde N_c} p_{N_f}(y)p_{\tilde N_c}(\tilde z^{-1}) \right. \cr
&\qquad\qquad\qquad\qquad + (pq)^{\frac 12 r_q}v^{-N_c/\tilde N_c}p_{N_f}(\tilde y)p_{\tilde N_c}(\tilde z^{-1}) - (pq)^{1 - \frac 12 r_q}v^{N_c/\tilde N_c}p_{N_f}(\tilde y^{-1})p_{\tilde N_c}(\tilde z) \cr
&\qquad\qquad\qquad\qquad + \left.\sum_{j=1}^\alpha \left((pq)^{r_Q + \frac 12 r_j}p_{N_f}(y)p_{N_f}(\tilde y^{-1}) - (pq)^{1 - r_Q - \frac 12 r_j}p_{N_f}(y^{-1})p_{N_f}(\tilde y) \right)\right)\,.
}} 
In the large $N$ limit, both \aie, \aim\ have the form \spdo, with
\eqn\af{
 f(p,q) = -\left(\frac{p}{1-p} + \frac{q}{1-q} - \frac{1}{(1-p)(1-q)}((pq)^{\frac 12 r_X} - (pq)^{1 - \frac 12 r_X}) \right)\,.
}
The equality of $f(p,q)$ for the two theories is due to the fact that $f(p,q)$ contains only information about the field content in the adjoint of the gauge group in each theory, and is independent of the rank of the gauge group. Thus, matching the electric and magnetic indices in the large $N$ limit does not automatically imply 't Hooft anomaly matching between the theories, although equality of the full index does. 

The quantities $g, \bar g, h$ of \spdo\ for the electric theory take the form
\eqn\aeln{\eqalign{
g_E(p,q,v,y,\tilde y) &= \frac{v}{(1-p)(1-q)}\left((pq)^{\frac 12 r_Q}p_{N_f}(y) - (pq)^{1 - \frac 12 r_Q}p_{N_f}(\tilde y)\right)\,,\cr
\bar g_E (p,q,v,y,\tilde y)&= \frac{v^{-1}}{(1-p)(1-q)}\left((pq)^{\frac 12 r_Q}p_{N_f}(\tilde y^{-1}) - (pq)^{1 - \frac 12 r_Q} p_{N_f}(y^{-1}) \right)\,,\cr
h_E(p,q,y,\tilde y) &= 0\,,
}}
while the magnetic analogs are 
\eqn\amln{\eqalign{
g_M(p,q,v,y,\tilde y) &= \frac{v^{N_c/\tilde N_c}}{(1-p)(1-q)}\left((pq)^{\frac 12 r_q}p_{N_f}(y^{-1}) - (pq)^{1 - \frac 12 r_q}p_{N_f}(\tilde y^{-1}) \right)\,, \cr
\bar g_M (p,q,v,y,\tilde y)&= \frac{v^{-N_c/\tilde  N_c}}{(1-p)(1-q)}\left((pq)^{\frac 12 r_q}p_{N_f}(\tilde y) - (pq)^{1 - \frac 12 r_q}p_{N_f}(y) \right)\,, \cr
h_M(p,q,y,\tilde y) &= \frac 1{(1-p)(1-q)}\sum_j\left((pq)^{r_Q + \frac 12 r_j}p_{N_f}(y)p_{N_f}(\tilde y^{-1}) \right. \cr 
&\qquad \qquad \qquad \qquad \left. - (pq)^{1 - r_Q -\frac 12 r_j}p_{N_f}(y^{-1})p_{N_f}(\tilde y) \right)\,.
}}

Agreement of the large $N$ indices of the electric and magnetic theories implies that
\eqn\match{
\frac{g_E\bar g_E - g_M \bar g_M}{1 -f} = h_M- h_E
}
as functions of the variables $(p,q,v,y,\tilde y)$. 
%Isolating the part of $h_M$ that carries a factor of $p_{N_f}(y)p_{N_f}(\tilde y^{-1})$ and replacing $pq$ with $t^2$ \defpq, \match\  implies that 
Plugging in \aeln, \amln\ and reverting back from the variables $p,q$ to $t,x$ \defpq, \match\ implies that
\eqn\asm{
 \sum_{j=1}^\alpha t^{r_j} 
 % = \frac{1 - t^{\frac{4\alpha}{k+1}}}{(1 - t^{\frac{2}{k+1}})(1 + t^{\frac{2k}{k+1}})}
  =  \frac{ 1 + t^{\frac 2{k+1}} + t^{\frac 4{k+1}} + \dots + t^{\frac{4\alpha -2}{k+1}}}{1 + t^{\frac{2k}{k+1}}}
  = \frac{1 + \tilde t + {\tilde t}^2 + \dots + {\tilde t}^{2\alpha-1}}{1 + {\tilde t}^k}\,
}
with ${\tilde t} = t^{\frac 2{k+1}}$.
%
% Change the letter. 
%
 For \asm\ to have a solution with a finite number of positive $r_j$, 
all the roots of the polynomial in the denominator on the RHS must be contained among those of the numerator. The roots of the denominator are $\exp((2n -1)\pi i /k)$ for $n \in {\bf Z^+} \leq k\,,$ while those of the numerator are the $2\alpha$'th roots of unity.
Therefore, the smallest $\alpha$ for which \asm\ is a polynomial of a finite degree is
\eqn\atr{ 
\alpha = k\,.}
Other values with this property are $\alpha_n = nk$ for integer $n$. Both here and in all subsequent examples we discuss, the lowest value $(n=1)$ gives the physical meson spectrum on the l.h.s. Other choices of $n$ give rise to meson spectra that are easily ruled out from anomaly matching considerations, which as mentioned previously, are not automatically included in the large $N$ index matching between the electric and magnetic theories.

From \asm, \atr\ we can read off that
\eqn\arm{
r_j = \frac{2(j-1)}{k+1} = (j-1)\,r_X\,, \quad j = 1, \dots, k
}
suggesting that the magnetic mesons have the form $\tilde Q\tilde X^{j-1} Q$, $j = 1, \dots k,$ in agreement with the known duality of \refs{\KutasovVE\KutasovNP-\KutasovSS}. 

At the next level, the operator $\tilde Q X^k Q$ is set to zero by the equations of motion following from \akw. In the index, it does not contribute because it cancels against a fermionic operator with the same quantum numbers. The details of this cancellation are discussed in Appendix B.

\newsec{$D$ series}
The $D_{k+2}$ theory \BrodieVX\ is obtained by adding to the $A_k$ one an additional adjoint chiral superfield $Y$, with the superpotential 
\eqn\dkw{\WW = \Tr X^{k+1} + \Tr XY^2\,.}
It contains the fields

\bigskip
\vbox{
$$\vbox{\offinterlineskip
\hrule height 1.1pt
\halign{&\vrule width 1.1pt#
&\strut\quad#\hfil\quad&
\vrule width 1.1pt#
&\strut\quad#\hfil\quad&
\vrule width 1.1pt#
&\strut\quad#\hfil\quad&
\vrule width 1.1pt#
&\strut\quad#\hfil\quad&
\vrule width 1.1pt#
&\strut\quad#\hfil\quad&
\vrule width 1.1pt#
&\strut\quad#\hfil\quad&
\vrule width 1.1pt#\cr
height3pt
&\omit&
&\omit&
&\omit&
&\omit&
&\omit&
&\omit&
\cr
&\hfil Field&
&\hfil $SU(N_c)$&
&\hfil $SU(N_f)$&
&\hfil $SU(N_f)$&
&\hfil $U(1)_B$&
&\hfil $U(1)_R$&
\cr
height3pt
&\omit&
&\omit&
&\omit&
&\omit&
&\omit&
&\omit&
\cr
\noalign{\hrule height 1.1pt}
height3pt
&\omit&
&\omit&
&\omit&
&\omit&
&\omit&
&\omit&
\cr
&\hfil $Q$&
%&\hfil $N_c$& 
&\hfil $f$&
&\hfil $f$&
&\hfil $1$&
&\hfil $1$&
&\hfil $1 - \frac 1{k+1}\frac{N_c}{N_f}$& 
\cr
height3pt
&\omit&
&\omit&
&\omit&
&\omit&
&\omit&
&\omit&
\cr
%\noalign{\hrule}
%height3pt
&\omit&
&\omit&
&\omit&
&\omit&
&\omit&
&\omit&
\cr
&\hfil $\tilde Q$&
%&\hfil $N_c$& 
&\hfil $\bar f$&
&\hfil $1$&
&\hfil $\bar f$&
&\hfil $-1$&
&\hfil $1 - \frac 1{k+1}\frac{N_c}{N_f}$& 
\cr
%\noalign{\hrule}
height3pt
&\omit&
&\omit&
&\omit&
&\omit&
&\omit&
&\omit&
\cr
&\hfil $V$&
%&\hfil $\bar N_c$&
&\hfil ${\rm adj.}$&
&\hfil $1$&
&\hfil $1$&
&\hfil $0$&
&\hfil $0$&
\cr
%\noalign{\hrule}
height3pt
&\omit&
&\omit&
&\omit&
&\omit&
&\omit&
&\omit&
\cr
&\hfil $X$&
%&\hfil $\bar N_c$&
&\hfil ${\rm adj.}$&
&\hfil $1$&
&\hfil $1$&
&\hfil $0$&
&\hfil $\frac 2{k+1}$&
\cr
%\noalign{\hrule}
height3pt
&\omit&
&\omit&
&\omit&
&\omit&
&\omit&
&\omit&
\cr
&\hfil $Y$&
%&\hfil $\bar N_c$&
&\hfil ${\rm adj.}$&
&\hfil $1$&
&\hfil $1$&
&\hfil $0$&
&\hfil $\frac k{k+1}$&
\cr
%\noalign{\hrule}
height3pt
&\omit&
&\omit&
&\omit&
&\omit&
&\omit&
&\omit&
\cr
}\hrule height 1.1pt
}
$$
}
\centerline{Table 3: The field content of the $D_{k+2}$ electric theory.} 
\bigskip

\noindent Its single-particle index \spi, in terms of $p,q$ \defpq\ and other variables used in defining the $A_k$ electric index around \aie, is 
\eqn\die{\eqalign{
& i_E(p,q,v, y, \tilde y, z)  \cr
&= -\left(\frac{p}{1-p} + \frac{q}{1-q} - \frac{1}{(1-p)(1-q)}((pq)^{\frac 12r_X} - (pq)^{1 - \frac 12 r_X}) \right. \cr
&\quad  - \left. \frac{1}{(1-p)(1-q)}((pq)^{\frac 12r_Y} - (pq)^{1 - \frac 12 r_Y}) \right)(p_{N_c}(z) p_{N_c}(z^{-1})-1) \cr
&\quad  + \frac{1}{(1-p)(1-q)}\left((pq)^{\frac 12 r_Q}vp_{N_f}(y)p_{N_c}(z) - (pq)^{1 - \frac 12 r_Q}v^{-1}p_{N_f}(y^{-1})p_{N_c}(z^{-1}) \right. \cr
&\qquad\qquad\qquad\qquad +  \left. (pq)^{\frac 12 r_Q}v^{-1}p_{N_f}(\tilde y^{-1}) p_{N_c}(z^{-1}) - (pq)^{1 - \frac 12 r_Q}v p_{N_f}(\tilde y)p_{N_c}(z) \right)\,,
}}
with $r_X, r_Y, r_Q$ the $U(1)_R$-charges of the respective fields in Table 3. 

A dual description of the theory was proposed in \BrodieVX. The magnetic theory is $\NN=1$ SQCD with gauge group $SU(3kN_f - N_c)$ and the same charged matter as in the electric one, coupled to $3k$ singlet mesons 
\eqn\dkm{
M_{lj} = \tilde Q X^{l-1}Y^{j-1} Q, \qquad  l = 1, \dots, k, \quad  j = 1,2,3}
 via the superpotential 
 \eqn\dkmw{
 \WW = \Tr \tilde X^{k+1} + \Tr \tilde X \tilde Y^2 + \sum_{\ell = 1}^k\sum_{j=1}^3 M_{\ell j}\tilde q \tilde X^{k-\ell} \tilde Y^{3 - j} q\,.
 }
 A new element in the $D_k$ theories with even $k$  is the appearance of a quantum constraint on the chiral ring, associated with the adjoint chiral superfields. To summarize the main idea (see \refs{\BrodieVX,\IntriligatorMI,\KutasovYQA} for more detailed discussions), the classical equations of motion following from the superpotential of the electric theory \dkw, $X^k = Y^2$ and $\{X,Y\} = 0$, allow for mesons of the form $\tilde Q\Theta_{lj} Q$ with $\Theta_{lj} = X^{l-1}Y^{j-1};\,\, l = 1, \dots, k,\,\,j = 1,2, \dots$ These mesons are expected to appear in the magnetic theory as singlets of the magnetic gauge group, $M_{lj}$. Such fields indeed appear, \dkm, but the index $j$ takes only three values, $j=1,2,3$. For odd $k$  this is OK, since the classical F-term equations imply the chiral ring constraint $Y^3 = 0$.  For even $k$  such a constraint does not appear at the classical level, but is believed to appear quantum mechanically.  Our goal in this section is to repeat the steps carried out in the previous one for the $A$ series, and see if large $N$ index matching leads to the meson spectrum appearing in the $D$ series duality. For even $k$, this provides a further check on the presence of the quantum constraint. 

Again, we assume that the electric theory has an unknown dual description with gauge group $SU(\tilde N_c) = SU(\alpha N_f - N_c)$ with $\alpha$ an unspecified integer, and the fields
\bigskip
\vbox{
$$\vbox{\offinterlineskip
\hrule height 1.1pt
\halign{&\vrule width 1.1pt#
&\strut\quad#\hfil\quad&
\vrule width 1.1pt#
&\strut\quad#\hfil\quad&
\vrule width 1.1pt#
&\strut\quad#\hfil\quad&
\vrule width 1.1pt#
&\strut\quad#\hfil\quad&
\vrule width 1.1pt#
&\strut\quad#\hfil\quad&
\vrule width 1.1pt#
&\strut\quad#\hfil\quad&
\vrule width 1.1pt#\cr
height3pt
&\omit&
&\omit&
&\omit&
&\omit&
&\omit&
&\omit&
\cr
&\hfil Field&
&\hfil $SU(\tilde N_c)$&
&\hfil $SU(N_f)$&
&\hfil $SU(N_f)$&
&\hfil $U(1)_B$&
&\hfil $U(1)_R$&
\cr
height3pt
&\omit&
&\omit&
&\omit&
&\omit&
&\omit&
&\omit&
\cr
\noalign{\hrule height 1.1pt}
height3pt
&\omit&
&\omit&
&\omit&
&\omit&
&\omit&
&\omit&
\cr
&\hfil $q$&
%&\hfil $N_c$& 
&\hfil $f$&
&\hfil $\bar f$&
&\hfil $1$&
&\hfil $N_c/\tilde{N_c}$&
&\hfil $1 - \frac 1{k+1}\frac{\tilde N_c}{N_f}$& 
\cr
height3pt
&\omit&
&\omit&
&\omit&
&\omit&
&\omit&
&\omit&
\cr
%\noalign{\hrule}
%height3pt
&\omit&
&\omit&
&\omit&
&\omit&
&\omit&
&\omit&
\cr
&\hfil $\tilde q$&
%&\hfil $N_c$& 
&\hfil $\bar f$&
&\hfil $1$&
&\hfil $f$&
&\hfil $-N_c/\tilde{N_c}$&
&\hfil $1 - \frac 1{k+1}\frac{\tilde N_c}{N_f}$& 
\cr
%\noalign{\hrule}
height3pt
&\omit&
&\omit&
&\omit&
&\omit&
&\omit&
&\omit&
\cr
&\hfil $\tilde V$&
%&\hfil $\bar N_c$&
&\hfil ${\rm adj.}$&
&\hfil $1$&
&\hfil $1$&
&\hfil $0$&
&\hfil $0$&
\cr
%\noalign{\hrule}
height3pt
&\omit&
&\omit&
&\omit&
&\omit&
&\omit&
&\omit&
\cr
&\hfil $\tilde X$&
%&\hfil $\bar N_c$&
&\hfil ${\rm adj.}$&
&\hfil $1$&
&\hfil $1$&
&\hfil $0$&
&\hfil $\frac 2{k+1}$&
\cr
%\noalign{\hrule}
height3pt
&\omit&
&\omit&
&\omit&
&\omit&
&\omit&
&\omit&
\cr
&\hfil $\tilde Y$&
%&\hfil $\bar N_c$&
&\hfil ${\rm adj.}$&
&\hfil $1$&
&\hfil $1$&
&\hfil $0$&
&\hfil $\frac k{k+1}$&
\cr
%\noalign{\hrule}
height3pt
&\omit&
&\omit&
&\omit&
&\omit&
&\omit&
&\omit&
\cr
&\hfil $M_j, \,\, j = 1, \dots \alpha$&
%&\hfil $\bar N_c$&
&\hfil $1$&
&\hfil $f$&
&\hfil $\bar f $&
&\hfil $0$&
&\hfil $2r_Q + r_j$&
\cr
%\noalign{\hrule}
height3pt
&\omit&
&\omit&
&\omit&
&\omit&
&\omit&
&\omit&
\cr
}\hrule height 1.1pt
}
$$
}
\centerline{Table 4: The ``conjectured" field content of the $D_{k+2}$ magnetic theory.} 
\bigskip
\noindent This leads to a magnetic single-particle index \spi\ with the same form as that of the $A$ series index \aim, up to a contribution from $\tilde Y$:
\eqn\dim{\eqalign{
&i_M(p,q,v,y,\tilde y, \tilde z) \cr
&= -\left(\frac{p}{1-p} + \frac{q}{1-q} - \frac{1}{(1-p)(1-q)}((pq)^{\frac 12r_X} - (pq)^{1 - \frac 12 r_X}) \right. \cr
&\quad  - \left. \frac{1}{(1-p)(1-q)}((pq)^{\frac 12r_Y} - (pq)^{1 - \frac 12 r_Y}) \right)(p_{\tilde N_c}(\tilde z) p_{\tilde N_c}(\tilde z^{-1})-1) \cr
&\quad + \frac{1}{(1-p)(1-q)}\left((pq)^{\frac 12 r_q} v^{N_c/\tilde N_c} p_{N_f}(y^{-1}) p_{\tilde N_c}(\tilde z) - (pq)^{1 - \frac 12 r_q} v^{-N_c/\tilde N_c} p_{N_f}(y)p_{\tilde N_c}(\tilde z^{-1}) \right. \cr
&\qquad\qquad\qquad\qquad + (pq)^{\frac 12 r_q}v^{-N_c/\tilde N_c}p_{N_f}(\tilde y)p_{\tilde N_c}(\tilde z^{-1}) - (pq)^{1 - \frac 12 r_q}v^{N_c/\tilde N_c}p_{N_f}(\tilde y^{-1})p_{\tilde N_c}(\tilde z) \cr
&\qquad\qquad\qquad\qquad + \left.\sum_{j=1}^\alpha \left((pq)^{r_Q + \frac 12 r_j}p_{N_f}(y)p_{N_f}(\tilde y^{-1}) - (pq)^{1 - r_Q - \frac 12 r_j}p_{N_f}(y^{-1})p_{N_f}(\tilde y) \right)\right)\,.
}}

 In the large $N$ limit, the electric and magnetic indices have the form \spdo\ with 
\eqn\df{\eqalign{
 f(p,q) &= -\left(\frac{p}{1-p} + \frac{q}{1-q} - \frac{1}{(1-p)(1-q)}((pq)^{\frac 12 r_X} - (pq)^{1 - \frac 12 r_X}) \right. \cr
&\qquad \left. - \frac{1}{(1-p)(1-q)}((pq)^{\frac 12 r_Y} - (pq)^{1 - \frac 12 r_Y}) \right)\,.
}}
The functions $g,\bar g, h$ in \spdo\ for both theories can easily be read off from \die, \dim. Plugging into \match, we find that the analog of \asm\ for the $D$ series is
\eqn\dsm{
\sum_{j=1}^\alpha t^{r_j} 
%= \frac{1 - t^{\frac{2\alpha}{k+1}}}{(1 - t^{\frac{2}{k+1}})} 
= \frac{ 1 + t^{\frac 2{k+1}} + t^{\frac 4{k+1}} + \dots + t^{\frac{2\alpha -2}{k+1}}}{1 + t^{\frac{2k}{k+1}}- t^{\frac{k}{k+1}}} = \frac{1 + {\tilde t}^2 + {\tilde t}^4 + \dots + {\tilde t}^{2(\alpha-1)}}{1 - {\tilde t}^{k} + {\tilde t}^{2k}}%(1 - t^{\frac{2\alpha}{k+1}}) 
\,,
}
with ${\tilde t} = t^{\frac 1{k+1}}.$ 
For \dsm\ to have a solution with a finite number of positive $r_j$,
the roots of the polynomial in the denominator on the RHS must be contained among those of the numerator. 
 The minimal value of $\alpha$ with this property is
\eqn\dtr{
\alpha = 3k\,.
}
As before, $\alpha = 3nk$ with $n =2,3,\cdots$ are also solutions to \dsm, but can be ruled out by anomaly matching.
The values of $r_j$ obtained from \dsm, \dtr\ are easily checked to be those of \dkm.

The formulae for the indices of the $D_{k+2}$ theory and its magnetic dual for finite $N$ were given in \SpiridonovZA. The equality of the electric and magnetic indices is a conjecture that awaits a proof.

One might hope to see the quantum constraint $Y^3 = 0$ explicitly in the index by expanding it to appropriate order in the fugacities. Unfortunately, the presence or absence of this constraint is obscured by the appearance of many operators at the same order as $Y^3$. We discuss the details in Appendix B.

\newsec{$E_7$}
The $E_7$ theory is again $\NN=1$ SQCD with two adjoint chiral superfields $X,Y$, but with the superpotential 
\eqn\esevenw{
\WW = \Tr Y^3 + \Tr YX^3\,.
}
This determines the R-charges of the fields to be those listed in Table 5. 
The corresponding single-particle index \spi\ is given by eq. \die\ but with $r_X,\, r_Y,\, r_Q$ taking the values from Table 5.  

\bigskip
\vbox{
$$\vbox{\offinterlineskip
\hrule height 1.1pt
\halign{&\vrule width 1.1pt#
&\strut\quad#\hfil\quad&
\vrule width 1.1pt#
&\strut\quad#\hfil\quad&
\vrule width 1.1pt#
&\strut\quad#\hfil\quad&
\vrule width 1.1pt#
&\strut\quad#\hfil\quad&
\vrule width 1.1pt#
&\strut\quad#\hfil\quad&
\vrule width 1.1pt#
&\strut\quad#\hfil\quad&
\vrule width 1.1pt#\cr
height3pt
&\omit&
&\omit&
&\omit&
&\omit&
&\omit&
&\omit&
\cr
&\hfil Field&
&\hfil $SU(N_c)$&
&\hfil $SU(N_f)$&
&\hfil $SU(N_f)$&
&\hfil $U(1)_B$&
&\hfil $U(1)_R$&
\cr
height3pt
&\omit&
&\omit&
&\omit&
&\omit&
&\omit&
&\omit&
\cr
\noalign{\hrule height 1.1pt}
height3pt
&\omit&
&\omit&
&\omit&
&\omit&
&\omit&
&\omit&
\cr
&\hfil $Q$&
%&\hfil $N_c$& 
&\hfil $f$&
&\hfil $f$&
&\hfil $1$&
&\hfil $1$&
&\hfil $1 - \frac 1{9}\frac{N_c}{N_f}$& 
\cr
height3pt
&\omit&
&\omit&
&\omit&
&\omit&
&\omit&
&\omit&
\cr
%\noalign{\hrule}
%height3pt
&\omit&
&\omit&
&\omit&
&\omit&
&\omit&
&\omit&
\cr
&\hfil $\tilde Q$&
%&\hfil $N_c$& 
&\hfil $\bar f$&
&\hfil $1$&
&\hfil $\bar f$&
&\hfil $-1$&
&\hfil $1 - \frac 1{9}\frac{N_c}{N_f}$& 
\cr
%\noalign{\hrule}
height3pt
&\omit&
&\omit&
&\omit&
&\omit&
&\omit&
&\omit&
\cr
&\hfil $V$&
%&\hfil $\bar N_c$&
&\hfil ${\rm adj.}$&
&\hfil $1$&
&\hfil $1$&
&\hfil $0$&
&\hfil $0$&
\cr
%\noalign{\hrule}
height3pt
&\omit&
&\omit&
&\omit&
&\omit&
&\omit&
&\omit&
\cr
&\hfil $X$&
%&\hfil $\bar N_c$&
&\hfil ${\rm adj.}$&
&\hfil $1$&
&\hfil $1$&
&\hfil $0$&
&\hfil $\frac 49$&
\cr
%\noalign{\hrule}
height3pt
&\omit&
&\omit&
&\omit&
&\omit&
&\omit&
&\omit&
\cr
&\hfil $Y$&
%&\hfil $\bar N_c$&
&\hfil ${\rm adj.}$&
&\hfil $1$&
&\hfil $1$&
&\hfil $0$&
&\hfil $\frac 23$&
\cr
%\noalign{\hrule}
height3pt
&\omit&
&\omit&
&\omit&
&\omit&
&\omit&
&\omit&
\cr
}\hrule height 1.1pt
}
$$
}
\centerline{Table 5: The field content of the $E_7$ electric theory.} 

\bigskip

In \KutasovYQA\ we proposed a magnetic dual description for this theory, that has gauge group $SU(30k N_f - N_c)$, coupled to thirty singlet mesons $M_j\leftrightarrow\tilde Q\Theta_j (X,Y)Q,\,\, j = 1, \dots, 30$ via a superpotential similar to \dkmw. The specific form of the $\Theta_j(X,Y)$'s as ordered products of $X,Y$ can be found in \KutasovYQA.
As with the $D_{k+2}$ theories with even $k$, the classical chiral ring is larger. In particular, the number of operators $\Theta_j$ that can be used to make chiral mesons  is larger than thirty (and depends on $N_c$). In \KutasovYQA\ we proposed a quantum constraint on the chiral ring of the electric theory, that truncates this classical set to the thirty operators compatible with the duality. To provide further evidence for the validity of this constraint, we would like to repeat the discussion of the $A$ and $D$ series for this case. 

Thus, we again assume the existence of a magnetic dual with gauge group $SU(\tilde N_c) = SU(\alpha N_f - N_c)$ for an unknown integer $\alpha$, and  the fields
\bigskip
\vbox{
$$\vbox{\offinterlineskip
\hrule height 1.1pt
\halign{&\vrule width 1.1pt#
&\strut\quad#\hfil\quad&
\vrule width 1.1pt#
&\strut\quad#\hfil\quad&
\vrule width 1.1pt#
&\strut\quad#\hfil\quad&
\vrule width 1.1pt#
&\strut\quad#\hfil\quad&
\vrule width 1.1pt#
&\strut\quad#\hfil\quad&
\vrule width 1.1pt#
&\strut\quad#\hfil\quad&
\vrule width 1.1pt#\cr
height3pt
&\omit&
&\omit&
&\omit&
&\omit&
&\omit&
&\omit&
\cr
&\hfil Field&
&\hfil $SU(\tilde N_c)$&
&\hfil $SU(N_f)$&
&\hfil $SU(N_f)$&
&\hfil $U(1)_B$&
&\hfil $U(1)_R$&
\cr
height3pt
&\omit&
&\omit&
&\omit&
&\omit&
&\omit&
&\omit&
\cr
\noalign{\hrule height 1.1pt}
height3pt
&\omit&
&\omit&
&\omit&
&\omit&
&\omit&
&\omit&
\cr
&\hfil $q$&
%&\hfil $N_c$& 
&\hfil $f$&
&\hfil $\bar f$&
&\hfil $1$&
&\hfil $N_c/\tilde{N_c}$&
&\hfil $1 - \frac 19\frac{\tilde N_c}{N_f}$& 
\cr
height3pt
&\omit&
&\omit&
&\omit&
&\omit&
&\omit&
&\omit&
\cr
%\noalign{\hrule}
%height3pt
&\omit&
&\omit&
&\omit&
&\omit&
&\omit&
&\omit&
\cr
&\hfil $\tilde q$&
%&\hfil $N_c$& 
&\hfil $\bar f$&
&\hfil $1$&
&\hfil $f$&
&\hfil $-N_c/\tilde{N_c}$&
&\hfil $1 - \frac 19\frac{\tilde N_c}{N_f}$& 
\cr
%\noalign{\hrule}
height3pt
&\omit&
&\omit&
&\omit&
&\omit&
&\omit&
&\omit&
\cr
&\hfil $\tilde V$&
%&\hfil $\bar N_c$&
&\hfil ${\rm adj.}$&
&\hfil $1$&
&\hfil $1$&
&\hfil $0$&
&\hfil $0$&
\cr
%\noalign{\hrule}
height3pt
&\omit&
&\omit&
&\omit&
&\omit&
&\omit&
&\omit&
\cr
&\hfil $\tilde X$&
%&\hfil $\bar N_c$&
&\hfil ${\rm adj.}$&
&\hfil $1$&
&\hfil $1$&
&\hfil $0$&
&\hfil $\frac 49 $&
\cr
%\noalign{\hrule}
height3pt
&\omit&
&\omit&
&\omit&
&\omit&
&\omit&
&\omit&
\cr
&\hfil $\tilde Y$&
%&\hfil $\bar N_c$&
&\hfil ${\rm adj.}$&
&\hfil $1$&
&\hfil $1$&
&\hfil $0$&
&\hfil $\frac  23$&
\cr
%\noalign{\hrule}
height3pt
&\omit&
&\omit&
&\omit&
&\omit&
&\omit&
&\omit&
\cr
&\hfil $M_j, \,\, j = 1, \dots \alpha$&
%&\hfil $\bar N_c$&
&\hfil $1$&
&\hfil $f$&
&\hfil $\bar f $&
&\hfil $0$&
&\hfil $2r_Q + r_j$&
\cr
%\noalign{\hrule}
height3pt
&\omit&
&\omit&
&\omit&
&\omit&
&\omit&
&\omit&
\cr
}\hrule height 1.1pt
}
$$
}
\centerline{Table 6: The ``conjectured" field content of the $E_7$ magnetic theory.} 
\bigskip
\noindent 
Here $r_j$ are the $U(1)_R$ charges of $\Theta_j$ and, as before, we do not place any constraints on them.

The single-letter index of the theory of Table 6 is given by \dim\ with the the appropriate R-charges. Taking the large $N$ limit, the analog of \asm\ obtained from reading off the $f,g,h$ functions \spdo\ of the electric and magnetic single-letter indices and using \match\ now reads
\eqn\esm{
\sum_{j=1}^\alpha t^{r_j} %= \frac{1 - t^{\frac \alpha9}}{(1 - t^{\frac 19})(1 + t^{\frac 19} - t^{\frac 13} - t^{\frac 49} - t^{\frac 59} + t^{\frac 79} + t^{\frac 89})} 
= \frac{1 + t^{\frac 2 9} + t^{\frac 49} + \dots + t^{\frac {2(\alpha -1)}9}}{1 + t^{\frac 29} - t^{\frac 23} - t^{\frac 89} - t^{\frac {10}9} + t^{\frac {14}9} + t^{\frac {16}9}} %(1 - t^{\frac{2\alpha}{k+1}}) 
\,\,.
}
\esm\ can again only be satisfied with finite $\alpha$ if every root of the denominator on the r.h.s. coincides with a root of the numerator, which are $\alpha$th roots of unity. However, it is easy to check that this is in fact the case when $\alpha = 30$.\foot{Again, we discard solutions with $\alpha = 30n$ for positive integer $n>1$.} The r.h.s. is then a sum of thirty terms of the form $t^{r_j}$, with the $r_j$ coinciding with the meson spectrum found in \KutasovYQA. This provides further support for the picture proposed in \KutasovYQA.

Expanding the index to the level of the constraint, one encounters again the same situation as in the $D$ series, as discussed in Appendix B.

In principle, one can go beyond the Veneziano large $N$ limit, and compare the indices of the electric and magnetic theories for all $N_f,N_c$. Following \refs{\DolanQI, \SpiridonovZA} and the building blocks described in Appendix A, the index of the electric $E_7$ theory is
\eqn\eeexact{\eqalign{
&I_E(p,q,y, \tilde y) = \frac{(p;p)^{{N_c}-1}(q;q)^{{N_c}-1}}{N_c!} \Gamma((pq)^{r_X/2},(pq)^{r_Y/2}; p,q)^{N_c-1} \cr
&\times \int_{\bf T^{{N_c}-1}} \prod_{j=1}^{{N_c}-1}\frac{dz_j}{2\pi i z_j}  
\prod_{i=1}^{N_f} \prod_{j=1}^{N_c} \Gamma((pq)^{r_Q/2}y_iz_j, (pq)^{r_Q/2}\tilde y_i^{-1}z_j^{-1}; p,q) \cr
 & \times\prod_{1 \leq i < j \leq N_c}\frac{\Gamma((pq)^{r_X/2} z_i z_j^{-1}, (pq)^{r_X/2}  z_i^{-1}z_j, (pq)^{r_Y/2}  z_iz_j^{-1}, (pq)^{r_Y/2}  z_i^{-1}z_j; p,q)}{\Gamma(z_iz_j^{-1}, z_i^{-1}z_j; p,q)} \cr
}}
where ${\bf T}$ is the unit circle, $y, \tilde y, z$ are (as before) complex eigenvalues for the matrix representations of $SU(N_f) \times SU(N_f) \times SU(N_c)$ \zzd, and we have used the short-hand notation for products of elliptic gamma functions (see Appendix A),
\eqn\gcond{
\Gamma(y_1, y_2, \dots, y_k; p,q) = \Gamma(y_1; p,q) \times \Gamma(y_2;p,q) \times \cdots \times  \Gamma(y_k; p,q).
}

%the $r_i$ are the R-charges of the corresponding fields in Table 5, and $\prod y_i = \prod\tilde y_i = \prod z_i = 1$. 
%

\noindent Similarly, the magnetic index is
\eqn\emexact{\eqalign{
&I_M(p,q,y,\tilde y) = \frac{(p;p)^{\tilde N_c -1}(q;q)^{\tilde N_c-1}}{\tilde N_c!} \Gamma((pq)^{r_X/2},(pq)^{r_Y/2}; p,q)^{\tilde N_c-1} \cr
&\times  \prod_{n=1}^{30}\prod_{i,j = 1}^{N_f}\Gamma((pq)^{r_Q + \frac{r_{M_n}}{2}}y_i\tilde y_j^{-1};p,q) \cr
&\times \int_{\bf T^{\tilde N_c-1}}\prod_{j-1}^{\tilde N_c-1}\frac{dz_j}{2\pi i z_j} \prod_{i=1}^{N_f}\prod_{j=1}^{\tilde N_c}\Gamma((pq)^{r_q/2}y_i^{-1}z_j, (pq)^{r_q/2}\tilde y_i z_j^{-1};p,q)  \cr
&\times \prod_{1 \leq i < j \leq \tilde N_c} \frac{\Gamma((pq)^{r_X/2}z_i z_j^{-1}, (pq)^{r_X/2}z_i^{-1}z_j, (pq)^{r_Y/2}z_iz_j^{-1}, (pq)^{r_Y/2}z_i^{-1}z_j; p,q)}{\Gamma(z_iz_j^{-1}, z_i^{-1}z_j; p,q)}
}}
where $\tilde z_i$ are the eigenvalues of $SU(\tilde N_c)$ and the thirty $r_{M_n}$ are the ones read off from \esm\ with $\alpha = 30$.

As a corollary of our duality \KutasovYQA, we conjecture
\eqn\hpeq{
I_E = I_M
}
between \eeexact\ and \emexact\ as a hypergeometric integral identity.

\newsec{$E_6$ and $E_8$}

The two remaining ADE fixed points \IntriligatorMI\ are the $E_6$ and $E_8$ theories, which are obtained by turning on the superpotentials
\eqn\esixeight{\eqalign{
\WW_{E_6} &= \Tr Y^3 + \Tr X^4\,, \cr
\WW_{E_8} &= \Tr Y^3 + \Tr X^5\,.
}}
As discussed in \refs{\IntriligatorMI,\KutasovYQA} many of the properties of the other theories are expected to be present in these cases as well. In particular, there is evidence that these theories do not have a vacuum below a certain critical number of flavors, although it is not known what that number is. Furthermore, the UV descriptions of these theories in terms of $SU(N_c)$ gauge theory seems to break down in the infrared for sufficiently small $N_f$. Thus, it is natural to expect that there is a weakly coupled dual description that makes these two phenomena manifest, like in the other cases. 

As mentioned in \KutasovYQA, an $\NN=1$ dual with the same properties as in the other cases does not seem to exist. In this section, we use the large $N$ superconformal index to confirm this conclusion using the tools of the previous sections. 

The field content and transformation properties under the symmetries  of the $E_6$ and $E_8$ theories are the following:
\bigskip
\vbox{
$$\vbox{\offinterlineskip
\hrule height 1.1pt
\halign{&\vrule width 1.1pt#
&\strut\quad#\hfil\quad&
\vrule width 1.1pt#
&\strut\quad#\hfil\quad&
\vrule width 1.1pt#
&\strut\quad#\hfil\quad&
\vrule width 1.1pt#
&\strut\quad#\hfil\quad&
\vrule width 1.1pt#
&\strut\quad#\hfil\quad&
\vrule width 1.1pt#
&\strut\quad#\hfil\quad&
\vrule width 1.1pt#\cr
height3pt
&\omit&
&\omit&
&\omit&
&\omit&
&\omit&
&\omit&
\cr
&\hfil Field&
&\hfil $SU(N_c)$&
&\hfil $SU(N_f)$&
&\hfil $SU(N_f)$&
&\hfil $U(1)_B$&
&\hfil $U(1)_R$&
\cr
height3pt
&\omit&
&\omit&
&\omit&
&\omit&
&\omit&
&\omit&
\cr
\noalign{\hrule height 1.1pt}
height3pt
&\omit&
&\omit&
&\omit&
&\omit&
&\omit&
&\omit&
\cr
&\hfil $Q$&
%&\hfil $N_c$& 
&\hfil $f$&
&\hfil $f$&
&\hfil $1$&
&\hfil $1$&
&\hfil $1 - \frac 1{6}\frac{N_c}{N_f}$& 
\cr
height3pt
&\omit&
&\omit&
&\omit&
&\omit&
&\omit&
&\omit&
\cr
%\noalign{\hrule}
%height3pt
&\omit&
&\omit&
&\omit&
&\omit&
&\omit&
&\omit&
\cr
&\hfil $\tilde Q$&
%&\hfil $N_c$& 
&\hfil $\bar f$&
&\hfil $1$&
&\hfil $\bar f$&
&\hfil $-1$&
&\hfil $1 - \frac 1{6}\frac{N_c}{N_f}$& 
\cr
%\noalign{\hrule}
height3pt
&\omit&
&\omit&
&\omit&
&\omit&
&\omit&
&\omit&
\cr
&\hfil $V$&
%&\hfil $\bar N_c$&
&\hfil ${\rm adj.}$&
&\hfil $1$&
&\hfil $1$&
&\hfil $0$&
&\hfil $0$&
\cr
%\noalign{\hrule}
height3pt
&\omit&
&\omit&
&\omit&
&\omit&
&\omit&
&\omit&
\cr
&\hfil $X$&
%&\hfil $\bar N_c$&
&\hfil ${\rm adj.}$&
&\hfil $1$&
&\hfil $1$&
&\hfil $0$&
&\hfil $\frac 12$&
\cr
%\noalign{\hrule}
height3pt
&\omit&
&\omit&
&\omit&
&\omit&
&\omit&
&\omit&
\cr
&\hfil $Y$&
%&\hfil $\bar N_c$&
&\hfil ${\rm adj.}$&
&\hfil $1$&
&\hfil $1$&
&\hfil $0$&
&\hfil $\frac 23$&
\cr
%\noalign{\hrule}
height3pt
&\omit&
&\omit&
&\omit&
&\omit&
&\omit&
&\omit&
\cr
}\hrule height 1.1pt
}
$$
}
\centerline{Table 7: The field content of the $E_6$ electric theory.} 
\bigskip

\bigskip
\vbox{
$$\vbox{\offinterlineskip
\hrule height 1.1pt
\halign{&\vrule width 1.1pt#
&\strut\quad#\hfil\quad&
\vrule width 1.1pt#
&\strut\quad#\hfil\quad&
\vrule width 1.1pt#
&\strut\quad#\hfil\quad&
\vrule width 1.1pt#
&\strut\quad#\hfil\quad&
\vrule width 1.1pt#
&\strut\quad#\hfil\quad&
\vrule width 1.1pt#
&\strut\quad#\hfil\quad&
\vrule width 1.1pt#\cr
height3pt
&\omit&
&\omit&
&\omit&
&\omit&
&\omit&
&\omit&
\cr
&\hfil Field&
&\hfil $SU(N_c)$&
&\hfil $SU(N_f)$&
&\hfil $SU(N_f)$&
&\hfil $U(1)_B$&
&\hfil $U(1)_R$&
\cr
height3pt
&\omit&
&\omit&
&\omit&
&\omit&
&\omit&
&\omit&
\cr
\noalign{\hrule height 1.1pt}
height3pt
&\omit&
&\omit&
&\omit&
&\omit&
&\omit&
&\omit&
\cr
&\hfil $Q$&
%&\hfil $N_c$& 
&\hfil $f$&
&\hfil $f$&
&\hfil $1$&
&\hfil $1$&
&\hfil $1 - \frac 1{15}\frac{N_c}{N_f}$& 
\cr
height3pt
&\omit&
&\omit&
&\omit&
&\omit&
&\omit&
&\omit&
\cr
%\noalign{\hrule}
%height3pt
&\omit&
&\omit&
&\omit&
&\omit&
&\omit&
&\omit&
\cr
&\hfil $\tilde Q$&
%&\hfil $N_c$& 
&\hfil $\bar f$&
&\hfil $1$&
&\hfil $\bar f$&
&\hfil $-1$&
&\hfil $1 - \frac 1{15}\frac{N_c}{N_f}$& 
\cr
%\noalign{\hrule}
height3pt
&\omit&
&\omit&
&\omit&
&\omit&
&\omit&
&\omit&
\cr
&\hfil $V$&
%&\hfil $\bar N_c$&
&\hfil ${\rm adj.}$&
&\hfil $1$&
&\hfil $1$&
&\hfil $0$&
&\hfil $0$&
\cr
%\noalign{\hrule}
height3pt
&\omit&
&\omit&
&\omit&
&\omit&
&\omit&
&\omit&
\cr
&\hfil $X$&
%&\hfil $\bar N_c$&
&\hfil ${\rm adj.}$&
&\hfil $1$&
&\hfil $1$&
&\hfil $0$&
&\hfil $\frac 25$&
\cr
%\noalign{\hrule}
height3pt
&\omit&
&\omit&
&\omit&
&\omit&
&\omit&
&\omit&
\cr
&\hfil $Y$&
%&\hfil $\bar N_c$&
&\hfil ${\rm adj.}$&
&\hfil $1$&
&\hfil $1$&
&\hfil $0$&
&\hfil $\frac 23$&
\cr
%\noalign{\hrule}
height3pt
&\omit&
&\omit&
&\omit&
&\omit&
&\omit&
&\omit&
\cr
}\hrule height 1.1pt
}
$$
}
\centerline{Table 8: The field content of the $E_8$ electric theory.} 
\bigskip

\noindent The single-letter indices of these theories are again given by eq. \die\ but with the charges in tables 7 and 8.
If we repeat the procedure of the previous sections to match the indices in the large $N$ limit to those of a conjectured magnetic dual, with gauge group $SU(\alpha N_f - N_c)$ for unknown $\alpha$, magnetic quarks $q, \tilde q$, adjoints $\tilde X, \tilde Y$ and $\alpha$ magnetic mesons,
the analogs of \asm\ specifying the magnetic meson spectrum turn out to be
\eqn\essm{\eqalign{
E_6 &: \sum_j t^{r_j} = \frac{1 + t^{\frac 13} + t^{\frac 23} + \dots + t^{\frac {\alpha-1}3} }{1 + t^{\frac 13} - t^{\frac 12} - t^{\frac 5{6}} - t^{\frac 7{6}} + t^{\frac 43} + t^{\frac 53}}\,, \cr
E_8 &: \sum_j t^{r_j} = \frac{1+ t^{\frac 2{15}} + t^{\frac 4 {15}} + \dots + t^{\frac{2(\alpha-1)}{15}} }{1 + t^{\frac 2{15}} + t^{\frac 4{15}} - t^{\frac 23} - t^{\frac 45} - t^{\frac {14}{15}} - t^{\frac {16}{15}} - t^{\frac 65} + t^{\frac 85} + t^{\frac{26}{15}} + t^{\frac{28}{15}}}\,\,.}}
Unlike in the earlier examples, the roots of the polynomials in the denominators on the r.h.s. for both $E_6$ and $E_8$ (in terms of $\tilde t = t^{1/6},\,\, \tilde t= t^{2/15}$ respectively) do not all lie on the unit circle. Instead, in both cases, there are two pairs of roots equidistant from the unit circle (with their product having modulus 1), while all the other roots lie on the unit circle (see figure 1). 
\ifig\loc{The roots of the polynominals in the denominators of eq. (5.2) for $E_6$ (left) and $E_8$ (right).}
{\epsfxsize4in\epsfbox{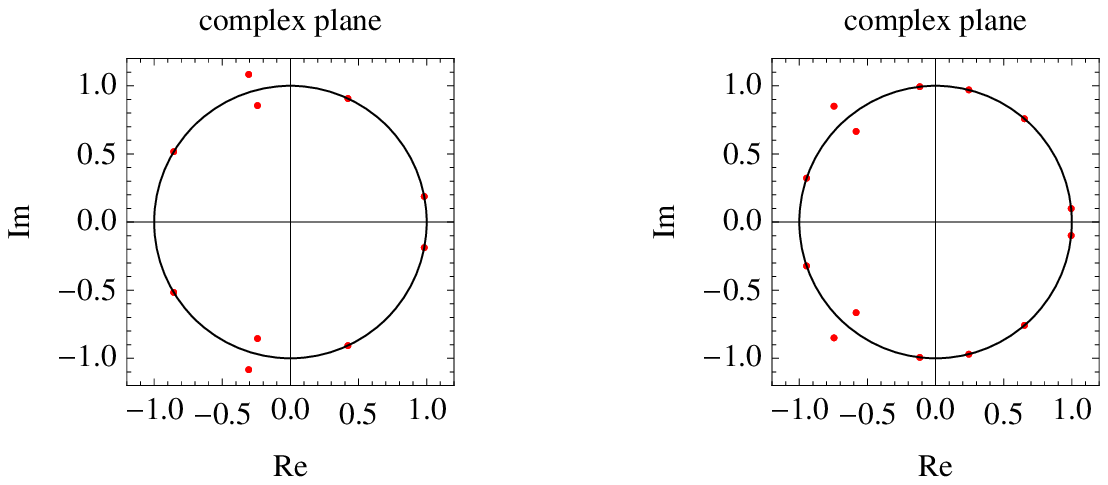}}

Therefore, the roots cannot all be contained among the roots of the numerators which are roots of unity. As discussed previously, this is a prerequisite for there to be a solution of \essm\ with finite $\alpha$. Therefore, if the $E_6$ and $E_8$ theories have dual descriptions, they must involve some new elements. It would be interesting to find them.

\bigskip 
 
\noindent{\bf Acknowledgements}: We thank O. Aharony for discussions. This work was supported in part by DOE grant DE-FG02-13ER41958, NSF Grant No. PHYS-1066293 and the hospitality of the Aspen Center for Physics, and by the BSF -- American-Israel Bi-National Science Foundation. The work of JL was supported in part by an NSF Graduate Research Fellowship. DK thanks IPMU Tokyo, Tel Aviv University and the Hebrew University for hospitality at various stages of this work.

\appendix{A}{Elliptic Hypergeometric Formulae for the Index}

In this section, we summarize the procedure for obtaining the exact superconformal index of a 4d $\NN=1$ theory (without going to the large $N$ limit) as a product of ``building blocks" for each multiplet \refs{\DolanQI,\SpiridonovZA}. 
The single-letter index \spi\ is a sum of contributions from each multiplet, so the full index \fulli\ which is the plethystic exponential of the single-letter one, can be decomposed as a product of contributions from individual multiplets.

In terms of the variables $p = tx, q = tx^{-1}$ \defpq\ and $y = t^r z$, where $y$ here has no relation to the fugacity for the global symmetry group appearing in the text, the single-letter index \spi\ for a chiral multiplet with R-charge $r$ is 
\eqn\isi{
i_S(p,q,y) = \frac{t^r z - t^{2-r}z^{-1}}{(1-tx)(1-tx^{-1})} = \frac{y - pq/y}{(1-p)(1-q)}\,.
}
Its contribution to the full index \fulli\ is  \DolanQI
\eqn\mpii{
\Gamma(y; p,q) = \exp\left(\sum_{n=1}^\infty \frac 1 n i_S(p^n, q^n, y^n) \right) = \prod_{j,k \geq 0} \frac{1 - y^{-1}p^{j+1}q^{k+1}}{1 - y p^jq^k}\,,
}
where $\Gamma(y;p,q)$ is the elliptic gamma function. 

For an abelian gauge field, the single-letter index is
\eqn\ivi{
i_V(p,q) =  -\frac{p}{1-p} - \frac{q}{1-q}
}
leading to the following contribution to the full index,
\eqn\mpvi{
\exp\left(\sum_{n=1}^\infty \frac 1n  i_V(p^n, q^n) \right) = (p;p)(q;q) }
where 
\eqn\sch{(x;p) = \prod_{j \geq 0} (1 - xp^j)\,.} 

Similar expressions are obtained for higher rank gange groups.
In particular when the gauge group is $SU(N)$, the contribution of the vector multiplet to the index together with that of the gauge group integral that projects out singlets \fulli\ yields \SpiridonovZA
\eqn\suni{
\frac{(p;p)^{N-1}(q;q)^{N-1}}{N!}\int_{\bf T^{N-1}}\prod_{i=1}^{N-1}\frac{dz_i}{2\pi i z_i}\prod_{1 \leq i < j \leq N}\frac{1}{\Gamma(z_iz_j^{-1}, z_i^{-1}z_j; p,q)}
}
where $\bf T$ is the unit circle and $z_i$ are the eigenvalues of $SU(N)$ \zzd. The index of a particular $SU(N)$ theory is the product of the building blocks \mpii, \suni, with one copy of \mpii\ for each chiral multiplet in the theory.

\appendix{B}{Structure of Low-Lying Terms}

In this section, we comment on the structure of low-lying terms in the $t$-expansion %(expansion in small $\RR$ \defr) 
of the superconformal index at large $N$, for the models discussed above. 
The hope is to explicitly see the quantum constraint in the $D_k$ for $k$ even and $E_7$ cases by checking that the coefficient of the index at the level where the constraint first appears is the right one for the truncated operator not to contribute. 
Unfortunately, we will find that the constraints in our examples appear at high enough levels in the expansion, that cancellation is obscured by the presence of a large number of operators that contribute at the same level.

We will make use of the fact that the $\NN=1$ index \fulli\ can be expanded as \DolanQI
\eqn\iex{
I(t,x,y) = \sum_{r,j,R_F} n_{r,j,R_F}\frac{t^r\chi_{2j+1}(x)}{(1-tx)(1-tx^{-1})}\chi_{R_F}(y)\,,
}
where $\chi_{2j+1}$ is the character for the representation of the $SU(2)$ in the isometry group of ${\bf S}^3$ with $J_3$ eigenvalue $j$ (see around \defr; all mentions of an $SU(2)$ in this appendix will refer to this one, as opposed to the other $SU(2)$ that has Cartan generator $\bar J_3$), 
 $\chi_{R_F}$ are characters of the representation $R_F$ of the flavor group $F$, and $n_{r,j,R_F}$ are integer coefficients counting the number of short multiplets in the $\NN=1$ SCFT that transform as $(r,j,R_F)$ under $(\RR,\, SU(2),\, F)$. 

A similar exercise was carried out for 3d $\NN=2$ Chern-Simons theories in \KapustinVZ, where quantum constraints in the chiral ring appear at low dimension, and the cancellation of truncated operators is visible in the $t$-expansion. See \BashkirovVY\ for a related discussion.

\subsec{$A$ series}
We first verify that in the $A$ series discussed in Section 2, the classical truncation of the chiral ring from the equations of motion for $X$ that follow from the superpotential \akw\ is visible in the expansion of the large $N$ formula \iln, \aie\ for the index. The e.o.m. for $X$ is 
\eqn\xeoma{
X^k - \frac 1 N (\Tr X^k)\,{\bf 1} = {\rm D\,\, term}\,,
}
hence the operators $\Tr X^l,\, l > k$ and those involving the matrix $X^l$ with $l \geq k$ are classically set to zero.   
%
%
%This is visible in the expansion of the index by extracting the relevant coefficient of $t$ in the large $N$ formula \iln, using \aie, \aeln. 

%(Note that the R and $\RR$ \defr\ charges of the components of chiral multiplets that we discuss in this section are the same because they have $\HH = 0$ and don't transform under $SU(2)_2$).
 %
In particular, the equation of motion implies that
\eqn\trunc{\tilde Q X^k Q  =0\,.}
Our goal is to see how \trunc\ is manifested in the index. 

The operator in \trunc\ transforms in the bifundamental of the flavor group $F = SU(N_f)_L \times SU(N_f)_R$, and is a singlet of $SU(2)$ with $j=0$. The relevant coefficient in \iex\ is thus associated to charges $r =  2r_Q + \frac{2k}{k+1}$, $j = 0$, and proportional to the flavor group character $p_{N_f}(y)p_{N_f}(\tilde y)$ \pnotn.

%--------------------------------------------------

We compute this coefficient in the expansion of the large $N$ formula \iln. 
Comparing with \aeln, in order to get a factor of $p_{N_f}(y)p_{N_f}(\tilde y)$ we keep terms that contain exactly one copy of $g_E\bar g_E$.
In other words, the number of chiral operators with $\RR$ charge $r + 2 r_Q$ that transform in the bifundamental of the flavor group are the coefficients of $t^r$ in the expansion of
% recall that pq is t^2, this is why there's a 1/2 in exponents in (2.6)...
%
\eqn\akone{
\frac{1}{1-f(t,x)}\prod_{n=1}^\infty \left[ \frac{1}{1-f(t^n,x^n)}\exp\left (-\frac 1 n f(t^n, x^n)\right)\right]\,,
}
with $f(t,x)$ given in \aeln.  Then to restrict to $j=0$, we further keep only terms in \akone\ with no explicit $x$ dependence.

%Focusing on the bifundamental operators that are also singlets of $SU(2)$ (with $j=0$ in (expi)),
Expanding \akone\ in powers of $t$, we find %$r - 2 R(Q) = 2n/(k+1) \leq 2k/(k+1)$ 
the following coefficients for $t^{2n/(k+1)}$ with integer $n < k$, along with the operators that they correspond to:

\eqn\coeff{
\matrix{
n & {\rm coefficient} & {\rm operators} \cr
1 & 1 & \tilde QXQ  \cr
2 & 2 & \tilde QX^2Q & \tilde QQ\Tr X^2 \cr
3 & 3 & \tilde QX^3Q & \tilde QXQ \Tr X^2 &  \tilde QQ \Tr X^3 \cr
4 & 5 & \tilde QX^4Q &  \tilde QX^2Q\Tr X^2 &  \tilde QXQ \Tr X^3 &\tilde QQ \Tr X^4 &  \tilde QQ(\Tr X^2)^2\cr
%5 & 7 & \tilde QX^5 Q, \tilde QX^3Q \Tr X^2, \tilde QX^2 Q \Tr X^3, \tilde QXQ \Tr X^4, \tilde QXQ(\Tr X^2)^2, \cr
%& & \tilde QQ \Tr X^5, \tilde QQ\Tr X^3 \Tr X^2 \cr
\vdots & \vdots & \vdots
}}
However, at level $n = k$ there is a modification to \coeff. 
The coefficient of \akone\ at $n=k$ is lower by 1 from the value listed in the table.
As a specific example, expanding \akone\ for $k=4$ and dropping terms with $x$ dependence, we find the polynomial $$1 + t^{2/5} + 2t^{4/5} + 3t^{6/5} + 4t^{8/5} + \dots$$ which agrees with \coeff\ for $n=1,2,3,$ but has coefficient 4 instead of 5 at level $n=4$.

This occurs because the conjugate fermion $\bar\psi_X$ in the $X$ multiplet, which contributes to the index with $\RR$ charge of  $2-r_X = 2 - \frac{2}{k+1} = \frac{2k}{k+1}$ as discussed around \spi, 
first appears at the $n=k$ level carrying a minus sign from $(-1)^F$. 
The index therefore says that the operator $\tilde Q X^k Q$ may combine with $\tilde Q\bar\psi_X Q$ to form a long multiplet.\foot{In principle, we know only that $\tilde Q\bar\psi_XQ$ cancels against {\it some} operator in the list \coeff. However, here it 
cannot cancel against anything except for \trunc, because all the other ones are multitrace operators that are products of single-trace ones appearing already at lower levels. $\tilde QX^kQ$ is the only new single-trace operator appearing at level $n=k$.} 
This is consistent with the fact that $\tilde QX^kQ$ is not in the chiral ring. 

In general, for any theory with a chiral multiplet $\phi$, the contribution of $\bar\psi_\phi$ is how the index implements the classical equation of motion for $\phi$. The F-term equation for $\phi$ has the same $\RR$-charge $2 - r_\phi$ as $\bar\psi_\phi$, which contributes to the index with opposite sign.

\subsec{$D$ series and $E_7$}
We now return to the question of whether the quantum constraint on the chiral ring in examples discussed above, $Y^3 = 0$ for the $D_{k+2}$ with $k$ even models \BrodieVX\ and a more complicated analog for the $E_7$ theory \KutasovYQA, can be seen by the index. 
The answer appears to be no for the following reason.
In the $D$ series, the operator $Y$ has R-charge $k/(k+1)$ (see Table 3), so the quantum constraint has charge 
\eqn\ryd{
R\,(Y^3) = \frac{3k }{k+1} \geq 2
}
for all $k$ even, with equality when $k=2$. Similarly, the $E_7$ quantum constraint has R-charge $30/9 > 2$ \KutasovYQA.

 However at $\RR=2$, $SU(2)$ singlet operators that involve the vector multiplet and/or spacetime derivatives, which have $\HH=0$ and are counted by the index, first begin to appear. 
 
 For example, looking at the bifundamental operators, in addition to those operators  built out of $\tilde Q, Q, X, \bar\psi_X$ that are listed in table \coeff, there are additional $SU(2)$ singlets such as 
$$
\tilde Q\lambda^2 Q,\,\, \tilde QQ\Tr\lambda^2,\,\, \partial_{-+}\tilde Q\lambda_- Q,\,\, \dots
$$
where $\lambda$ is the gaugino in the vector multiplet, and in the derivative $\partial_{\alpha\dot \alpha}$ the first index is that of the $SU(2)$ discussed throughout this appendix, while the second index is for the other $SU(2)$ with generator $\bar J_3$.\foot{This is not a complete list. See \RomelsbergerEC\ for such a list in the case of ordinary Seiberg duality.} $\lambda$ and $\partial_{\pm +}$ both carry $\RR$-charges 1 \defr, so these operators first appear starting at $\RR = 2$. (Two or more $\lambda$'s and $\partial_{\pm +}$'s are required, to contract spinor indices and get an $SU(2)$ singlet). 

Given the appearance of many single-trace bosonic and ferminoic operators which contribute to the index at levels $\RR \geq 2$, there are multiple cancellations among them, which obscure the appearance or absence of the quantum constraint.

%\appendix{C}{Group Characters}
%
%In this section we briefly discuss the properties of group characters and their appearance in the index.  
%
%For a group $G$ with a representation $R$, the character $\chi_{R}: G \rightarrow {\bf C}$ is defined s.t. $\chi_R(U)$ is the trace of the group element $U$ in the representation $R$.
%
%
%(*) 

\listrefs

\bye